\documentclass[11pt,a4paper]{article}

\usepackage{amsmath}
\usepackage{amssymb}
\usepackage{graphicx}
\usepackage[margin=2.6cm]{geometry}
\usepackage[font=small]{caption}

\newcommand{\Ig}{I_{\mathrm{G}}}
\newcommand{\cmi}{I(n_1\!\colon\! n_3\,|\,n_2)}
\newcommand{\Ex}{\mathbb{E}}

\title{An information bound for multiplicities in rapidity windows\\ of the dipole cascade behind the entanglement entropy picture}
\author{
  Olasantan Ebenezer Adelaja \qquad Alex Prygarin\thanks{Corresponding author.} \qquad
  Karam Shekh Yusuf\\[6pt]
  \small Department of Physics, Ariel University, Ariel 40700, Israel\\[2pt]
  \small \texttt{olasanta.adelaja@msmail.ariel.ac.il}\\[1pt]
  \small \texttt{alexanderp@ariel.ac.il}\\[1pt]
  \small \texttt{yusufk@ariel.ac.il}}
\date{}

\newcommand{\NumBoundShort}{0.1438}  % XS6/numbers_extra.txt, copied by name
\newcommand{\NumCmiGamma}{0.04031}  % XS6/numbers.txt, copied by name
\newcommand{\NumCmiShortEight}{0.1250}  % XS6/numbers_extra.txt, copied by name
\newcommand{\NumCmiShortFour}{0.0944}  % XS6/numbers_extra.txt, copied by name
\newcommand{\NumCmiShortZero}{0.0379}  % XS6/numbers_extra.txt, copied by name
\newcommand{\NumCmiUnif}{0.05455}  % XS6/numbers.txt, copied by name
\newcommand{\NumCompExcessTwo}{5.8}  % XS6/numbers_direct.txt, copied by name
\newcommand{\NumCompShortCutsOne}{5.5}  % XS6/numbers_direct.txt, copied by name
\newcommand{\NumFigCountOne}{2}  % SHORT_figs/data/numbers_fig1.txt, copied by name
\newcommand{\NumFigCountThree}{7}  % SHORT_figs/data/numbers_fig1.txt, copied by name
\newcommand{\NumFigCountTwo}{5}  % SHORT_figs/data/numbers_fig1.txt, copied by name
\newcommand{\NumFigRate}{0.33}  % XS6/numbers_extra.txt, copied by name
\newcommand{\NumFigTwoCritMid}{1.036}  % XS6/numbers.txt, copied by name
\newcommand{\NumFigTwoCrossMid}{1.305}  % XS6/numbers.txt, copied by name
\newcommand{\NumHOneCells}{sixteen}  % XS6/numbers_more.txt, copied by name
\newcommand{\NumIgUnif}{0.04258}  % XS6/numbers.txt, copied by name
\newcommand{\NumIgYeight}{0.1290}  % XS6/numbers.txt, copied by name
\newcommand{\NumIgYfour}{0.1028}  % XS6/numbers.txt, copied by name
\newcommand{\NumIgYzero}{0.0447}  % XS6/numbers.txt, copied by name
\newcommand{\NumLHCbRatioOne}{0.824}  % XS6/numbers_extra.txt, copied by name
\newcommand{\NumLHCbRatioThree}{0.992}  % XS6/numbers_extra.txt, copied by name
\newcommand{\NumLHCbRatioTwo}{0.886}  % XS6/numbers_extra.txt, copied by name
\newcommand{\NumMIuncondEight}{1.0009}  % XS6/numbers_extra.txt, copied by name
\newcommand{\NumMIuncondFour}{0.3935}  % XS6/numbers_extra.txt, copied by name
\newcommand{\NumMIuncondZero}{0.0770}  % XS6/numbers_extra.txt, copied by name
\newcommand{\NumNormTol}{2\times10^{-11}}  % XS6/numbers.txt, copied by name
\newcommand{\NumRatioShortHi}{0.97}  % XS6/numbers_extra.txt, copied by name
\newcommand{\NumRatioShortLo}{0.85}  % XS6/numbers_extra.txt, copied by name
\newcommand{\NsDeepLimit}{0.1374}  % closed form at Delta = 0.4, tau = 1, checked against XS6/numbers*.txt NumDeepLimit
\newcommand{\NsOccHi}{7.1}  % calc_SHORT/item3/dep_crossing.log, 100 a<N_R>, unit windows from y0 = 4, k = 1
\newcommand{\NsOccLo}{5.3}  % calc_SHORT/item3/dep_crossing.log, 100 a<N_R>, unit windows from y0 = 4, k = 3
\newcommand{\NsSizeNullHi}{4.9}  % calc_SHORT/item2/power_results.json, pooled rate over the null cells, largest of order2, order3, gamma, bonf
\newcommand{\NsSizeNullLo}{4.7}  % calc_SHORT/item2/power_results.json, pooled rate over the null cells, smallest of order2, order3, gamma, bonf
\newcommand{\NsSizeSe}{0.18}  % calc_SHORT/item2/power_results.json, sqrt(p(1-p)/R) of the pooled null rates, largest, rounded up
\newcommand{\NsStarSigma}{6.9}  % (0.58 - 1/6)/0.06, Bzdak 1108.0882 after Eq. (10) and footnote 3
\newcommand{\NsStarSigmaZdc}{5.6}  % (0.5 - 1/6)/0.06, Bzdak 1108.0882 "very close to 1/2" and footnote 3
\newcommand{\NsTwoPointExcess}{0.0053}  % calc_SHORT/item2/exact_info.json, table1 twopoint, I - I_G, checked by the chain rule here
\newcommand{\NsUnifCutsDiff}{0.0008}  % calc_SHORT/item5/map_sign.json, cuts k=3 uniform[0,2], -(I - I_G)
\newcommand{\NtAbove}{14}  % calc_SHORT/item13/summary.json sign_pc '>', counted over the triples
\newcommand{\NtAboveRatio}{4.0}  % calc_SHORT/item13/summary.json, largest I_corr / IG_pc among the triples above ('A19', '0', 3)
\newcommand{\NtBgHi}{79}  % calc_SHORT/item12, 100 (1 - sqrt F), largest cell
\newcommand{\NtBgLo}{48}  % calc_SHORT/item12, 100 (1 - sqrt F), smallest cell
\newcommand{\NtDrawsAbove}{15}  % uvcheck/verify and v2, draws of ('U14', 1, 0) (undecided in the analysis) that lie above
\newcommand{\NtDrawsFurther}{27}  % uvcheck/verify (18 draws) and verify/v2 (9 draws) per triple, counted
\newcommand{\NtDrawsUndec}{5}  % uvcheck/verify and v2, draws of ('U14', 1, 1) (above in the analysis) that are undecided
\newcommand{\NtEffHi}{57}  % calc_SHORT/item12b/occupancy_observed_mean.json u, largest reachable cell
\newcommand{\NtEffLo}{33}  % calc_SHORT/item12b/occupancy_observed_mean.json u, smallest reachable cell
\newcommand{\NtExcessSigma}{eleven}  % calc_SHORT/item13/summary.json within.z and verify/results, smallest of both 11.71 rounded down
\newcommand{\NtFMax}{0.266}  % calc_SHORT/item12/h1_implied_k.json 2021 as printed, largest F of 16 cells
\newcommand{\NtFMin}{0.044}  % calc_SHORT/item12/h1_implied_k.json 2021 as printed, smallest F of 16 cells
\newcommand{\NtNoCapTriples}{42}  % calc_SHORT/item13/summary.json, triples of A14 and A19 (no cap), counted
\newcommand{\NtOccUnreach}{seven}  % calc_SHORT/item12b, cells with H1 F below F_min, counted
\newcommand{\NtPzeroHi}{7.5}  % H1 2021 Table 9-12 P(0), percent, largest cell
\newcommand{\NtPzeroLo}{0.2}  % H1 2021 Table 9-12 P(0), percent, smallest cell
\newcommand{\NtQcmiZero}{0.1558}  % calc_SHORT/item15/quantum_cmi.json, y0 = 0, closed form, checked against the direct route
\newcommand{\NtTriples}{84}  % calc_SHORT/item13/summary.json, triples of A14, A14cap, A19, A19cap counted
\newcommand{\NtZHi}{43.7}  % calc_SHORT/item12, (1 - F)/sF as printed, largest
\newcommand{\NtZLo}{11.5}  % calc_SHORT/item12, (1 - F)/sF as printed, smallest

\begin{document}
\maketitle

\begin{abstract}
\noindent
In Mueller's dipole cascade without transverse dimensions, with one counted particle for each
dipole produced in a window, the multiplicities in three consecutive rapidity windows are Poisson
counts of one Gamma-distributed source. We show that at fixed middle multiplicity the outer two share
less information than the Gaussian value $-\tfrac12\ln(1-\rho^2)$ of their partial correlation, a
value that second moments alone determine. For windows of one common width at a constant splitting
rate this value never exceeds $\tfrac12\ln(4/3)$. The Gaussian value is not a bound on mutual
information in general, and for a source of the same mean and variance with another law the
information can exceed it. The law of the counts keeps its form under migration of particles across
window edges and detection losses when both act on each particle independently of the others and of
the source, and its factorial
cumulants of second and third order test necessary conditions for it. In Monte Carlo simulations of the cascade with transverse dimensions at leading logarithmic
accuracy and fixed coupling, the counts are not Poisson counts of one source, so the bound is a result
of the model without them. In the model without transverse dimensions the law also fixes the
normalized second factorial cumulant of one window at $1/k$, with $k$ the number of initial dipoles.
The values of this cumulant formed from the mean and variance that H1 publishes for deep inelastic
scattering lie far below the value one of a cascade from a single initial dipole with one counted
particle for each dipole produced in a window.
The quantity bounded is a classical conditional mutual information between counted multiplicities.
\end{abstract}

\section{Introduction}

The entanglement entropy of the proton is discussed through the multiplicity distributions of
particles produced in restricted rapidity intervals~\cite{KL,HKKT}, and in the dipole cascade behind
this picture the particles in different intervals descend from one branching history. Let $n_1$,
$n_2$ and $n_3$ be the numbers of particles counted in three consecutive rapidity windows on the
same events. The middle window measures the overall activity of the cascade, so holding $n_2$ fixed
removes much of what the outer counts have in common. The outer counts still share information at
fixed $n_2$.

In Mueller's dipole cascade~\cite{Mueller1994,Mueller1995} with the transverse dimensions
removed~\cite{Mueller1995,KL}, and with one counted particle for each dipole produced in a window,
the three counts are Poisson counts of one Gamma-distributed source. For these counts the information
that remains is bounded by a number that depends on second moments alone,
\begin{equation}
  \cmi\;<\;-\tfrac12\ln\!\left(1-\rho^2\right)\;\le\;\tfrac12\ln\tfrac43 ,
  \label{eq:main}
\end{equation}
where $\rho$ is the correlation of the outer counts at fixed middle count, equal in this cascade to
their partial correlation, the correlation of $n_1$ and $n_3$ after the linear dependence of both on
$n_2$ is removed, and measured from two regression slopes. The left-hand side, called the information left in what follows, is the
classical conditional mutual information of $n_1$ and $n_3$ given $n_2$. It measures how much knowing
one outer count tells about the other once the middle count is known, and it vanishes precisely when the two
are independent at every value of the middle count that occurs. The information left would
equal the middle term if the two counts were jointly Gaussian with correlation $\rho$, and this term
is called the Gaussian value. The right-hand side is called the ceiling. The first inequality holds
wherever the three windows lie in rapidity and at every width, splitting rate and number of initial
dipoles. For the second it is enough that the mean counts $\lambda_j$ of the windows satisfy
$\lambda_1\lambda_3\le\lambda_2^2$, which three consecutive windows of one common width at a
constant splitting rate satisfy with equality.

The Gaussian value is not a bound on mutual information in general. Kraskov, St\"ogbauer and
Grassberger stated it as a lower bound at given correlation~\cite{KSG2004} and withdrew the proof in
an erratum that restricts the statement to Gaussian marginals~\cite{KSGerratum}. Foster and
Grassberger showed that such bounds depend on the marginal distributions and derived them for
Gaussian and for uniform marginals~\cite{FosterGrassberger}. Cardoso gave a case, credited there to
Plumbley, in which the Gaussian value exceeds the information~\cite{Cardoso2003}, and Pires and
Perdig\~ao a discrete one~\cite{PiresPerdigao}. The direction of the inequality therefore depends on
the joint law of the counts. Equation~(\ref{eq:main}) fixes it for the law the cascade produces, and
for some other sources with the same mean and variance, a uniform one among them, the information
left exceeds the Gaussian value in some window configurations.

The second inequality of Eq.~(\ref{eq:main}) is equivalent to $|\rho|\le\tfrac12$, and an upper limit
of one half on a correlation at fixed multiplicity has been obtained before under other hypotheses. Lappi and McLerran~\cite{LappiMcLerran} bound the partial correlation of two
pseudorapidity intervals at fixed multiplicity in a wider third interval between them by one half in
a Gaussian approximation. Bzdak~\cite{Bzdak} obtains the same limit in the same geometry for a linear
regression of an outer count on the count in the third interval. Both derivations assume a
correlation function that depends on the two pseudorapidities only through their separation and does
not increase with it. Olszewski and
Broniowski~\cite{OlszewskiBroniowski} take the partial covariance with respect to a central control
bin. For two regions filled by a binomial split of the total, Biyajima, Blazek and
Suzuki~\cite{Biyajima1989} give the moments of one count at fixed value of the other. For a
pure-birth process started from a Poisson number of particles they find the mean of one count at fixed
value of the other linear in the fixed count only asymptotically.

Carruthers and Shih~\cite{CS1989} evaluated the mutual information
between two rapidity regions from forward-backward multiplicity data and wrote down a three-region
measure. Kharzeev~\cite{Kharzeev2026} proposes the mutual information of the counts in two jet
regions as a probe of quantum entanglement in jet fragmentation. Equation~(\ref{eq:main}) bounds the
information left. It follows from the law of the three counts,
which we call the window law. The window law also sets the normalized second factorial cumulant
within every window and between any two windows to one common value. Since the second inequality
needs only the window means, an information left above $\tfrac12\ln(4/3)$ in windows with
$\lambda_1\lambda_3\le\lambda_2^2$ excludes the window law wherever the windows lie, at every
splitting rate and number of initial dipoles.

The window law keeps its form under the independent migration and loss of the counted particles, with
new window means, and its factorial cumulants of second and third order test necessary conditions for it. The bound belongs to the model without
transverse dimensions, since in Monte Carlo simulations of the cascade with them at leading logarithmic accuracy
and fixed coupling, the counts are no longer Poisson counts of one source. Published data bear on the
window law as well. At the mean counts that LHCb publishes for its pseudorapidity intervals, each mean taken
over the distribution truncated at twenty particles~\cite{LHCbmult}, the window law satisfies the condition for the ceiling. In central gold-gold collisions the correlation that STAR reports at
fixed multiplicity in a reference interval~\cite{STARfb,Bzdak} lies $\NsStarSigma$ standard deviations
above the limit $1/6$ that the window law sets for the intervals of that measurement. The window law
fixes the normalized second factorial cumulant of one window at the inverse of the number of initial
dipoles, and in the H1 data on deep inelastic scattering~\cite{H1mult} this cumulant lies far below the
value one of a single initial dipole.

The structure of this paper is as follows. Section~\ref{sec:law} gives the window law, its
representation by one Gamma source and its factorial cumulants, and Sec.~\ref{sec:bound} derives
Eq.~(\ref{eq:main}). Section~\ref{sec:robust} shows which features of hadronization and detection
leave the window law intact and how it is tested, Sec.~\ref{sec:beyond} turns to the cascade with
transverse dimensions, and Sec.~\ref{sec:data} to published data. Section~\ref{sec:interpret} relates
the quantity bounded to the entanglement entropy picture.

\section{The window law}\label{sec:law}

In the limit of a large number $N_c$ of colors and at leading logarithmic accuracy, Mueller wrote the
light-cone wave function of a heavy quark-antiquark state as a classical branching process. A soft
gluon counts in color as a quark-antiquark pair, so each emission splits one pair into two, with a
probability that depends on their transverse separations~\cite{Mueller1994}. In a later paper, which
calls these pairs color dipoles, Mueller introduces a model without transverse dimensions, whose quanta,
defined in analogy with the dipoles, split into two at a constant rate, so that the number of quanta
from one parent is geometric~\cite{Mueller1995}. Levin
and Lublinsky write the master equation of this model~\cite{LevinLublinsky}, and
Kharzeev and Levin, who describe its quanta as dipoles of a fixed size, use it for the entanglement
entropy and call it a toy $(1{+}1)$ dimensional model~\cite{KL}. We call it the $1{+}0$ model, with one
dimension in rapidity and none transverse, as Hentschinski, Kutak, P{\l}aczek and Rohrmoser
do~\cite{HKPR}.

Let one dipole split at the constant rate $\Delta$ in the rapidity $y=\ln(1/x)$, with $x$ the
longitudinal momentum fraction, and start the cascade from $k$ dipoles at $y=0$. We call a fixed
rapidity interval $[y_{j-1},y_j]$ a window and $n_j$ the number of splittings in it. Each splitting
raises the number of dipoles by one, so $n_j$ is also the number of dipoles produced in the window.
Nesting the one-dipole generating function over consecutive windows gives the joint law of the counts
in three windows~\cite{BatesNeyman},
\begin{equation}
  P(n_1,n_2,n_3)=\frac{\Gamma(k+N)}{\Gamma(k)\,n_1!\,n_2!\,n_3!}\;
  a_1^{n_1}a_2^{n_2}a_3^{n_3}\,(1-a_1-a_2-a_3)^{k},
  \qquad N=n_1+n_2+n_3,
  \label{eq:nm}
\end{equation}
the negative multinomial of shape $k$~\cite{Sibuya}, with $a_j=\lambda_j/(k+\Lambda)$,
$\Lambda=\lambda_1+\lambda_2+\lambda_3$ and window means
\begin{equation}
  \lambda_j=k\left(e^{\Delta y_j}-e^{\Delta y_{j-1}}\right).
  \label{eq:means}
\end{equation}
Equation~(\ref{eq:nm}) is the window law. Each window taken alone is negative binomial of shape $k$
and mean $\lambda_j$, geometric at $k=1$. For a window that starts at the origin of the cascade, this
negative binomial is the solution, with $k=2h$, of the cascade equation that Caputa and Kutak write with
a parameter $h$~\cite{CaputaKutak}. Kutak and L\"ok\"os identify $2h$ with the shape parameter of the
negative binomial~\cite{KutakLokos}. Equation~(\ref{eq:nm}) is a probability law at every real $k>0$,
and no step below uses an integer $k$, so Eq.~(\ref{eq:nm}) and the results derived from it hold at
every real $k>0$. At non-integer $k$ they describe a pure-birth process whose rate after $n$
productions is $\Delta(k+n)$, and $k$ is the number of initial dipoles when it is an integer. A marginal of Eq.~(\ref{eq:nm}) over any subset of windows is
again negative multinomial with the same shape~\cite{Sibuya}, so the region below the first window may
be left unobserved. A splitting rate that depends on rapidity leaves Eq.~(\ref{eq:nm}) unchanged, with
$\Delta y_j$ in Eq.~(\ref{eq:means}) replaced by the integral of the rate up to $y_j$.

\begin{figure}[t]
  \centering
  \includegraphics[width=0.92\textwidth]{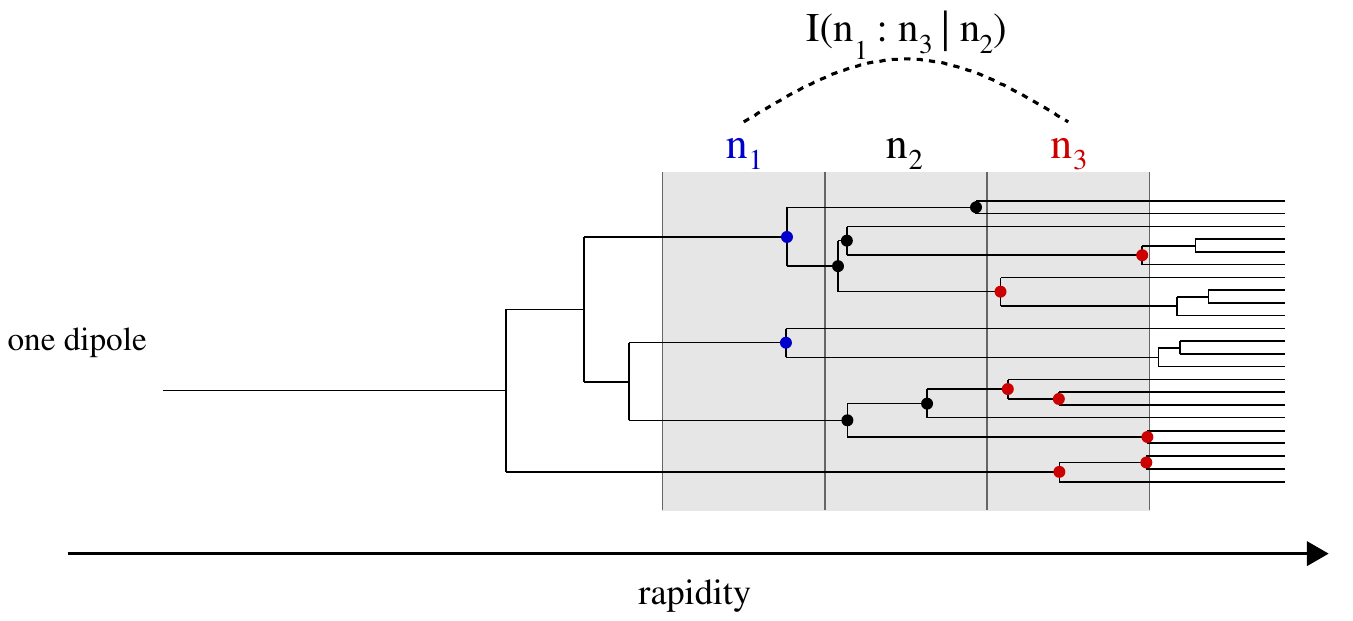}
  \caption{One realization of the $1{+}0$ model at the constant splitting rate
  $\Delta=\NumFigRate$, drawn as a branching history in rapidity from a single dipole at the left.
  The gray band is divided by two vertical lines into three consecutive rapidity windows and the
  marked points are the splittings inside them, so that $n_1=\NumFigCountOne$, $n_2=\NumFigCountTwo$ and
  $n_3=\NumFigCountThree$ here. The arc marks the information left between the outer two
  counts at fixed middle count.}
  \label{fig:history}
\end{figure}

Equation~(\ref{eq:nm}) is the law of Poisson counts driven by one fluctuating source. Given a
positive random variable $W$ of unit mean,
\begin{equation}
  n_j\,|\,W\;\sim\;\mathrm{Poisson}(\lambda_jW),\qquad W\;\sim\;\mathrm{Gamma}\ \text{of unit mean and shape } k ,
  \label{eq:mixture}
\end{equation}
independently in the three windows, and integrating over $W$ returns Eq.~(\ref{eq:nm})~\cite{Sibuya}.
The cascade
realizes this form event by event. The rapidities at which dipoles are produced form a Poisson
process whose intensity is $W$ times $k\Delta e^{\Delta y}$, with the same $W$ for the whole event.
In the variable $s=e^{\Delta y}-1$ it is, given $W$, a homogeneous Poisson process of rate $kW$, and
after the average over $W$ a P\'olya process. Bates and Neyman~\cite{BatesNeyman} give the generating
function over consecutive intervals and Kendall the characteristic functional~\cite{Kendall1952}. The
representation of the birth times of a linear pure-birth process as a mixed Poisson process is
credited by Pitman and Yakubovich to a later paper of Kendall~\cite{PitmanYakubovich}, and Polito
states it for a process started from one individual~\cite{Polito}. Appendix~\ref{app:proof} gives the
form for $k$ initial dipoles and a rapidity-dependent rate. The variable $W$ is the overall activity
of the cascade, and all correlation between the windows comes from it. Figure~\ref{fig:history} shows
one history.

The law factorizes into the negative binomial of the total count $N$, with shape $k$ and mean
$\Lambda$, times a multinomial split of $N$ among the windows with probabilities $\lambda_j/\Lambda$,
and the split does not contain $k$~\cite{Sibuya}. Since the shape is $k$
and every $\lambda_j$ is proportional to $k$, the correlation between windows is fixed by the window
edges and the splitting rate alone.

The factorial cumulants of the counts follow from Eq.~(\ref{eq:mixture}) for a source of any law. For
one common source with Poisson counting, the generating function obeys
$\ln\Ex\prod_j(1+t_j)^{n_j}=K_W(\sum_j\lambda_jt_j)$ with $K_W$ the cumulant generating function of
$W$. Every factorial cumulant of order $m$, divided by the product of the means it involves, is
therefore the $m$-th cumulant $\kappa_m(W)$ of $W$, the same for every choice of windows. This
equality is the single-source case of the relation between factorial cumulants of counts and
cumulants of the fluctuating densities reviewed by De Wolf, Dremin and Kittel~\cite{DeWolfDreminKittel}. At second
order the normalized second factorial cumulants are
\begin{equation}
  F_{jl}=\frac{\Ex[n_jn_l]-\delta_{jl}\lambda_j}{\lambda_j\lambda_l}-1=\mathrm{Var}\,W
  \label{eq:fcum}
\end{equation}
within one window and between any two. The Gamma source adds a relation between orders,
$\kappa_3=2\kappa_2^2$, which for three distinct windows reads
\begin{equation}
  \kappa_{123}=\frac{2\,C_{12}\,C_{23}}{\lambda_2} ,
  \label{eq:gamma3}
\end{equation}
with $\kappa_{123}$ the joint third cumulant and $C_{jl}$ the covariances. The negative multinomial
of Eq.~(\ref{eq:nm}) is the law with $\kappa_m=(m-1)!\,(\mathrm{Var}\,W)^{m-1}$ at every order. These
identities are those of Sibuya, Yoshimura and Shimizu~\cite{Sibuya} (Table~1, cumulant generating
function) and Van Hove~\cite{VanHove1987}, and for two windows they follow from the joint generating
function of Carruthers and Shih~\cite{CS1985}. At $k$ initial dipoles every $F_{jl}$ equals $1/k$.

\section{The bound on the information left}\label{sec:bound}

The information left is the classical mutual information of $n_1$ and $n_3$ at fixed $n_2$, averaged over
$n_2$. Here $b$ is a possible middle count, and conditioning on $b$ means conditioning on $n_2=b$,
\begin{equation}
  \cmi=\sum_{b\ge0}P(n_2=b)\,I(n_1\!:\!n_3\,|\,n_2=b) ,
  \label{eq:cmiavg}
\end{equation}
with the mutual information at $n_2=b$
\begin{equation}
  I(n_1\!:\!n_3\,|\,n_2=b)=\sum_{n_1,n_3}P(n_1,n_3\,|\,b)\,
  \ln\frac{P(n_1,n_3\,|\,b)}{P(n_1\,|\,b)\,P(n_3\,|\,b)} .
  \label{eq:cmib}
\end{equation}
For each $b$, the mutual information in Eq.~(\ref{eq:cmib}) is
$H(n_1|b)+H(n_3|b)-H(n_1,n_3|b)$, with $H$ the Shannon entropy of the distribution named. Averaging over
$n_2$ gives entropies conditioned on $n_2$, and the chain rule
$H(x\,|\,n_2)=H(x,n_2)-H(n_2)$ then gives the average in terms of joint entropies,
\begin{equation}
  \cmi=H(n_1,n_2)+H(n_2,n_3)-H(n_1,n_2,n_3)-H(n_2) ,
  \label{eq:cmidef}
\end{equation}
where $H(n_1,n_2)=-\sum_{n_1,n_2}P(n_1,n_2)\ln P(n_1,n_2)$ and the other entropies are formed in the
same way from the distribution of the counts named. All logarithms are natural, so the information is
given in nats. At fixed $n_2=b$ the outer counts obey a two-window negative multinomial, derived below, whose mutual
information is the $M(k+b)$ of Eq.~(\ref{eq:mixture2}), so that Eq.~(\ref{eq:mixture2}) is
Eq.~(\ref{eq:cmiavg}) for the window law. Without conditioning on the
middle window, the mutual information of the outer two counts is
\begin{equation}
  I(n_1\!:\!n_3)=H(n_1)+H(n_3)-H(n_1,n_3) .
  \label{eq:midef}
\end{equation}
The information left is what the outer counts still share once the middle count is known, and
$I(n_1\!:\!n_3)$ is the full information they share. No inequality between the two holds for every joint
law. For the window law at the three depths below, the information left is the smaller, as the values quoted
there show.
The information left vanishes precisely when $n_1$, $n_2$ and $n_3$ form a Markov chain, that is when
$n_1$ and $n_3$ are independent at fixed $n_2$, the middle count screening the outer two~\cite{HaydenJPW}. The Markov property of the cascade does not make the
middle window screen its neighbors. The population present at rapidity $y$ follows a linear
pure-birth process, and at fixed level of this population at an intermediate rapidity the earlier
population and the later one are independent. An experiment counts what is produced inside a window, not that level, and
knowing $n_2$ does not tell how many dipoles were present when the middle window began, so the outer
counts remain dependent.

The splitting rate $\Delta=0.4$ is the value of $4\bar{\alpha}_s\ln2$ at
$\alpha_s=0.15$ with $\bar{\alpha}_s=N_c\alpha_s/\pi$, rounded to one decimal~\cite{CaputaKutak}. At
this rate and from one dipole, the information left for three windows of unit width is
$\NumCmiShortZero$, $\NumCmiShortFour$ and $\NumCmiShortEight$ at $y_0=0$,
$4$ and $8$, where $y_0$, the depth of the windows, is the lower edge of the first window. The
information left grows as the windows are placed deeper, toward $-\tfrac12\ln[1-r/(1+r)^2]$, with
$r=e^{\Delta\tau}$ for windows of width $\tau$, which is $\NsDeepLimit$ for the windows of unit width. For the same unit windows, the mutual information
$I(n_1\!:\!n_3)$ of Eq.~(\ref{eq:midef}) grows without limit as $y_0$ increases. Its values are
$\NumMIuncondZero$, $\NumMIuncondFour$ and $\NumMIuncondEight$ at
the same depths. Every value at a finite depth is a sum over the counts weighted by the window law,
Eq.~(\ref{eq:nm}), and the probabilities kept in each sum add up to one within $\NumNormTol$.

An observed middle count $n_2=b$ changes what is known about the source. The posterior law of $W$ is
a Gamma of shape $k+b$, and the outer counts are again Poisson variables mixed over one Gamma. Their
law is the two-window negative multinomial of shape
\begin{equation}
  K=k+b,\qquad\text{means } Kc_j,\qquad c_j=\frac{\lambda_j}{k+\lambda_2} ,
  \label{eq:cj}
\end{equation}
the one-element case of Eq.~(2.6) of Sibuya, Yoshimura and Shimizu~\cite{Sibuya}, and for two windows
the relation of Carruthers and Shih~\cite{CS1985}. The mean of an outer count at fixed middle count
is linear,
\begin{equation}
  \Ex[n_j\,|\,n_2=b]=c_j\,k+c_j\,b ,
  \label{eq:reg}
\end{equation}
so the ratios $c_j$ are regression slopes that an experiment measures without knowing $k$ or
$\Delta$, and the ratio of intercept to slope is $k$~\cite{CS1985,Sibuya}. For two regions of equal
mean the slope is the forward-backward strength that Giovannini and Ugoccioni give for a
negative-binomial total split binomially between two hemispheres~\cite{GiovanniniUgoccioni}. The correlation of the outer counts at fixed
$n_2$ does not depend on $b$,
\begin{equation}
  \rho^2=\frac{c_1c_3}{(1+c_1)(1+c_3)} ,
  \label{eq:rho}
\end{equation}
and it equals the partial correlation of $n_1$ and $n_3$ given $n_2$ formed from the covariance
matrix $C$ of the three counts,
$(C_{13}-C_{12}C_{23}/C_{22})/\sqrt{(C_{11}-C_{12}^2/C_{22})(C_{33}-C_{23}^2/C_{22})}$, as Baba and Sibuya
prove for this family~\cite{BabaSibuya}.

The information left is an average over the middle count of a one-parameter family,
\begin{equation}
  \cmi=\sum_{b\ge0}P(n_2=b)\,M(k+b),
  \label{eq:mixture2}
\end{equation}
where $M(K)$ is the mutual information of the two-window negative multinomial of shape $K$ and ratios
$c_1$, $c_3$. As $K$ grows the law becomes Gaussian with correlation $\rho$~\cite{GenestOuimet}, and
$M(K)$ tends to the Gaussian value
\begin{equation}
  \Ig=\tfrac12\ln\frac{(1+c_1)(1+c_3)}{1+c_1+c_3}=-\tfrac12\ln\!\left(1-\rho^2\right).
  \label{eq:IG}
\end{equation}
This limit bounds every member of the family from above. Written in the form of
Eq.~(\ref{eq:mixture}), the two outer counts are Poisson of means $c_jKW'$ given a Gamma variable
$W'$ of unit mean and shape $K$, and they are independent given $W'$. For two counts that are independent
given $W'$, the mutual information of the counts is
$I(n_1\!:\!W')+I(n_3\!:\!W')-I(n_1,n_3\!:\!W')$, with $I(x\!:\!W')$ the mutual information of $x$ and
$W'$. At fixed total $n_1+n_3$ their split is binomial and does not involve $W'$, so the total is a
sufficient statistic for $W'$, which means that it carries all the information the pair holds about $W'$,
and $I(n_1,n_3\!:\!W')$ is the mutual information of the total and $W'$. The total is Poisson of mean
$(c_1+c_3)KW'$ given $W'$, and therefore
\begin{equation}
  M(K)=J(K,c_1)+J(K,c_3)-J(K,c_1+c_3) ,
  \label{eq:sep}
\end{equation}
where $J(K,c)$ is the mutual information between $W'$ and a count $n$ that is Poisson of mean $cKW'$
given $W'$. After the average over $W'$ the count $n$ is negative binomial of shape $K$ and mean $cK$. The
alternating form is the co-information of the two counts and the source, introduced by McGill in the
opposite sign convention~\cite{McGill1954}, and the binomial split is Eq.~(2.7) of Sibuya, Yoshimura
and Shimizu~\cite{Sibuya}. The recursion of the negative binomial gives
\begin{equation}
  \frac{\partial^2 J}{\partial c^2}=-\frac{\Ex\!\left[\chi(K+n)\right]}{(1+c)^2},
  \qquad \chi(z)=z(z+1)\ln\!\left(1+\tfrac1z\right)-z ,
  \label{eq:Jpp}
\end{equation}
and $\chi$ rises toward $\tfrac12$ and stays below it (both in Appendix~\ref{app:proof}). The function
$\Theta(c)=J(K,c)-\tfrac12\ln(1+c)$ vanishes at $c=0$, where the count is zero in every event, and by
Eq.~(\ref{eq:Jpp}) it has $\Theta''=[\tfrac12-\Ex\chi(K+n)]/(1+c)^2>0$, so it is strictly convex and
hence strictly superadditive, $\Theta(c_1)+\Theta(c_3)<\Theta(c_1+c_3)$ for $c_1,c_3>0$. Inserting it into
Eq.~(\ref{eq:sep}) gives
\begin{equation}
  M(K)=\left[\Theta(c_1)+\Theta(c_3)-\Theta(c_1+c_3)\right]+\Ig\;<\;\Ig
  \label{eq:proof}
\end{equation}
at every shape, and by Eq.~(\ref{eq:mixture2}) the information left obeys the same inequality. The
Gamma source enters this step through the recursion of the negative binomial, from which
Eq.~(\ref{eq:Jpp}) follows, and through the linear regression of Eq.~(\ref{eq:reg}), which makes the
Gaussian value the same for every member of the average. The inequality $\chi<\tfrac12$ that makes $\Theta$
convex is a property of the function $\chi$ alone. The other side, $\chi>0$, makes $J$ concave in
$c$, and this concavity holds for a non-negative source of any law, as Atar and Weissman
prove~\cite{AtarWeissman}. Martinez gives $J$ itself as a one-dimensional
integral~\cite{Martinez2008,MartinezThesis}.

The information left approaches the Gaussian value from below as the mean count of the middle window
grows. At fixed $c$ the
negative binomial is stochastically increasing in its shape~\cite{KlenkeMattner} and $\chi$ rises, so
for every $\delta>0$ Eq.~(\ref{eq:Jpp}) makes $g(c)=J(K+\delta,c)-J(K,c)$ strictly concave in $c$. This
difference vanishes at $c=0$ and is therefore strictly subadditive, $g(c_1)+g(c_3)>g(c_1+c_3)$, and
Eq.~(\ref{eq:sep}) gives
$M(K+\delta)>M(K)$. The difference $\Ig-M(K)$ falls as $\Xi/K$ at large $K$, with
$\Xi=\tfrac1{12}[(c_1+c_3)^2/(1+c_1+c_3)-c_1^2/(1+c_1)-c_3^2/(1+c_3)]$
(Appendix~\ref{app:proof}). Since $K=k+n_2$, the inequality $\cmi<\Ig$ is tight where the middle
window is well populated. At the three depths above, the Gaussian value is $\NumIgYzero$,
$\NumIgYfour$ and $\NumIgYeight$, and the ratio of the information left to it rises from $\NumRatioShortLo$ to
$\NumRatioShortHi$ between $y_0=0$ and $y_0=8$.

\begin{figure}[t]
  \centering
  \includegraphics[width=0.72\textwidth]{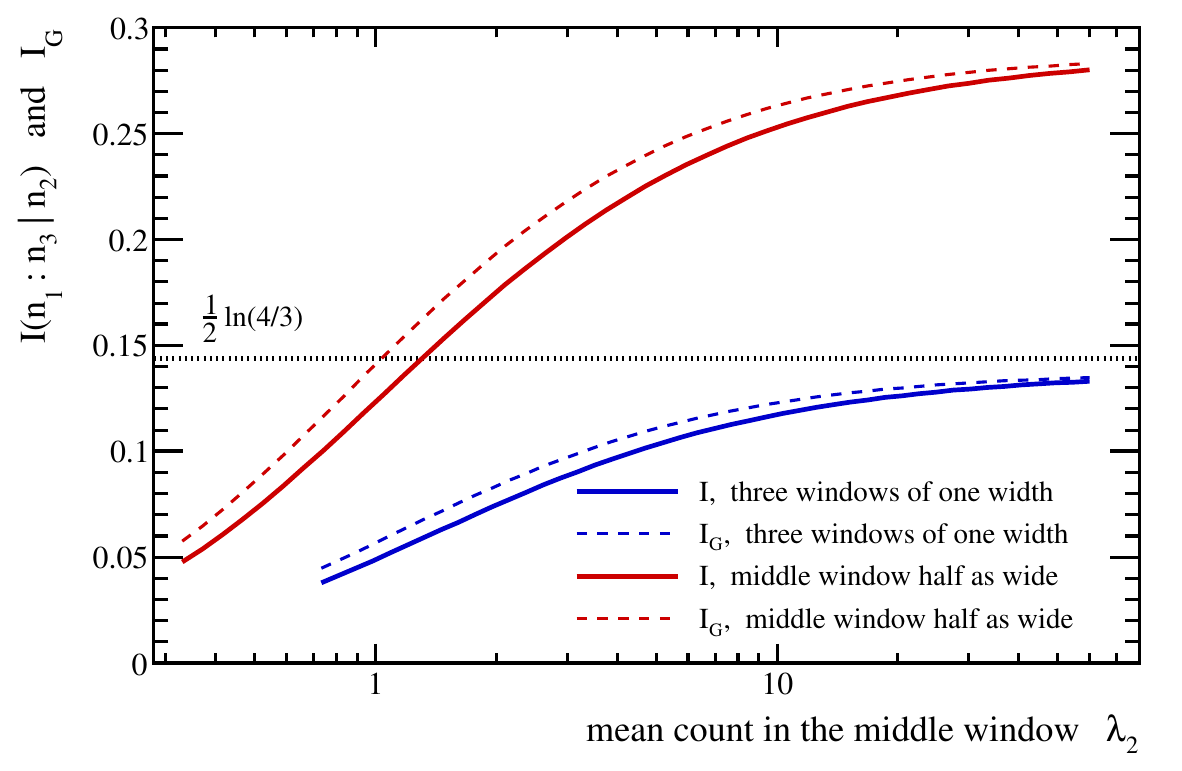}
  \caption{The information $\cmi$ left between the outer counts at fixed middle count (solid lines)
  and the Gaussian value $\Ig$ of Eq.~(\ref{eq:IG}) (dashed lines), against the mean count $\lambda_2$
  of the middle window, from sums of Eq.~(\ref{eq:nm}) for a cascade from one dipole at
  $\Delta=0.4$. Along each curve the three windows are moved deeper at fixed widths. Blue lines are
  three windows of unit width. Red lines are outer windows of unit width with a middle window half as
  wide, a geometry with $\lambda_1\lambda_3>\lambda_2^2$. The dotted line is $\tfrac12\ln(4/3)$. The
  information lies below $\Ig$ at every point. For the narrow middle window, $\Ig$ crosses
  $\tfrac12\ln(4/3)$ at $\lambda_2=\NumFigTwoCritMid$ and the information at
  $\lambda_2=\NumFigTwoCrossMid$. All logarithms are natural.}
  \label{fig:bound}
\end{figure}

For windows of one common width at a constant splitting rate the Gaussian value stays below the
ceiling. From
Eq.~(\ref{eq:IG}), $\Ig\le\tfrac12\ln(4/3)$ holds precisely when
\begin{equation}
  3c_1c_3\;\le\;1+c_1+c_3 ,
  \label{eq:sharp}
\end{equation}
which is satisfied whenever $\lambda_1\lambda_3\le\lambda_2^2$ at any number of initial dipoles
(Appendix~\ref{app:proof}). At a constant splitting rate three consecutive windows of width $\tau$ have
$\lambda_j=k\,e^{\Delta y_0}r^{\,j-1}(r-1)$ with $r=e^{\Delta\tau}$, so
\begin{equation}
  \lambda_1\lambda_3=\lambda_2^2
  \qquad\text{and}\qquad
  \cmi\;<\;\tfrac12\ln\tfrac43\;\simeq\;\NumBoundShort ,\quad |\rho|<\tfrac12 ,
  \label{eq:bound}
\end{equation}
at every depth, width, splitting rate and number of initial dipoles. Over this range the ceiling is
the least upper bound of both $\Ig$ and the information left, approached as the windows narrow and the
middle mean grows. The value $\tfrac12\ln(4/3)$ is the Gaussian value at $\rho=\tfrac12$, which is the
correlation limit of Lappi and McLerran and of Bzdak~\cite{LappiMcLerran,Bzdak}. The partial correlation formed from the covariance matrix
of the three counts depends on the law of the source only through $\mathrm{Var}\,W$. For Poisson
counts of any single source the square of this partial correlation is the right-hand side of Eq.~(\ref{eq:rho}), with $k$ in the
ratios $c_j$ of Eq.~(\ref{eq:cj}) replaced by $1/\mathrm{Var}\,W$. This partial correlation therefore
stays below $\tfrac12$ in absolute value, and the Gaussian value formed from it stays below
$\tfrac12\ln(4/3)$, for every such source whenever $\lambda_1\lambda_3\le\lambda_2^2$. The proof of the first inequality of
Eq.~(\ref{eq:main}) uses the Gamma source, and so does the equality of this partial correlation with
the correlation at fixed $n_2$. For a window geometry that violates $\lambda_1\lambda_3\le\lambda_2^2$, the Gaussian value and
the information itself can exceed $\tfrac12\ln(4/3)$, and Fig.~\ref{fig:bound} shows one geometry of
each kind. For windows of one common width the Gaussian value can also exceed it when the splitting
rate varies with rapidity or the efficiency differs between windows, since either changes the ratio
$\lambda_1\lambda_3/\lambda_2^2$.

The inequality $\cmi<\Ig$ holds for the Gamma source with Poisson counting, and it can fail for a
source with the same first two moments and another law. For a source uniform on $[0,2]$, which has
the variance of a Gamma of shape $3$, and three windows of equal mean $2$, the information left is
$\NumCmiUnif$, above $\Ig=\NumIgUnif$, while the Gamma source gives $\NumCmiGamma$.
For the same uniform source the information lies $\NsUnifCutsDiff$ below $\Ig$ for unit windows from
$y_0=4$ with $k=3$. For one and the same source law the information can therefore lie on either side
of $\Ig$, depending on the windows. For source laws matched in mean and variance, neither the third
cumulant relative to its Gamma value nor the weight of the upper tail determines the sign of $\cmi-\Ig$ at every configuration. For every Poisson-counted single source of
positive variance and finite third moment, the ratio of the information to $\Ig$ tends to
$2\varphi(v)/v^2$ as the three window means tend to zero in fixed ratios, where $v$ is the variance of
the source and $\varphi(v)=(1+v)\ln(1+v)-v$ (Appendix~\ref{app:proof}). This limit is below one, so the information lies below
$\Ig$ once the window means are small enough, and how small they must be depends on the source. Counting each dipole as
more than one particle also changes the information. Let each dipole give one particle plus a further
number of particles that is geometric with mean one, all counted in the window where the dipole was
produced. With the Gaussian value formed from the covariance matrix of the particle counts, the
information then lies $\NumCompExcessTwo$ percent above it for three windows of equal dipole mean $2$
with the Gamma source of shape $3$, and $\NumCompShortCutsOne$ percent below it for the unit windows
from $y_0=4$ with one initial dipole. A violation of the first inequality of Eq.~(\ref{eq:main})
therefore excludes the Gamma source with Poisson counting and does not show whether the source law or
the counting fails.

\section{Hadronization, detection and tests}\label{sec:robust}

An experiment counts hadrons, and a hadron need not land in the window where its dipole was
produced. Given $W$, the production rapidities are a Poisson process. If each dipole yields at most
one counted hadron, displaced in rapidity independently of all others by any kernel that does not
depend on $W$ and may depend on rapidity, the hadron rapidities are again a Poisson process given the same $W$, by the marking and
mapping theorems for Poisson processes~\cite{LastPenrose}. If each hadron is kept independently with
an efficiency that does not depend on $W$ and may depend on rapidity, the thinning theorem gives the same
conclusion~\cite{LastPenrose}. In both cases the counts in fixed windows remain the negative
multinomial of Eq.~(\ref{eq:nm}) with the same shape $k$, and only the window means change. Sibuya,
Yoshimura and Shimizu give the count-level form of the thinning statement~\cite{Sibuya}, and Van Hove
and Giovannini give the hadronization analogue with the shape unchanged under local parton-hadron
duality~\cite{VanHoveGiovannini}.

The observation model of this paper is therefore one hadron per dipole produced, with independent
migration and independent efficiency, and the ratios of Eq.~(\ref{eq:cj}) are formed from the observed
means. Hentschinski, Kutak, P{\l}aczek and Rohrmoser count one hadron per dipole in the total dipole
number, with an independent binomial thinning to charged hadrons~\cite{HKPR}. From here on $n_j$
denotes the number of hadrons counted in window $j$ and $\lambda_j$ its mean. In the comparisons with
data we identify the pseudorapidity of a hadron with the rapidity $y$ of its dipole, up to the
independent migration above, a change of sign and an additive constant common to the events of one
kinematic cell. The pseudorapidity is taken in the center-of-mass frame of the collision, which in
deep inelastic scattering is the hadronic center-of-mass frame. For the unit windows from $y_0=4$ at
$\Delta=0.4$, a Gaussian migration of width $\sigma$ raises the window means to about those of the
three windows placed $\Delta\sigma^2/2$ deeper. After such a migration three consecutive windows of
one common width $\tau$ still satisfy $\lambda_1\lambda_3\le\lambda_2^2$, and the ceiling of
Eq.~(\ref{eq:bound}) still applies. The mean count of the window $[y_0,y_0+\tau]$ is the integral over
the production rapidity $y'$ and the hadron rapidity $y$ of the density of dipole production,
$k\Delta e^{\Delta y'}$ for $y'>0$ and zero below, times the Gaussian of $y-y'$ and the indicator of
$y_0\le y\le y_0+\tau$. Each factor is log-concave in $(y',y,y_0)$, that is, its logarithm is concave there. Once the production density is
cut off at a large rapidity the product is integrable, and its integral over $y'$ and $y$ is then
log-concave in $y_0$~\cite{SaumardWellner}. The window mean is the limit of these integrals as the
cutoff grows, so it is log-concave in $y_0$ as well, which gives $\lambda_1\lambda_3\le\lambda_2^2$.

The window law is broken by clusters of hadrons from one dipole, by an efficiency that depends on the
occupancy of the event and by a displacement common to all hadrons of an event. If the dipoles yield
clusters of two or more hadrons, independently of one another and with sizes that do not depend on
$W$, and each hadron is displaced in rapidity independently of the others, the clusters add a
positive term to the second factorial cumulants. For displacements small compared with the
window width this term is largest within a window, smaller between adjacent windows and small between
windows that do not touch. An efficiency that falls with the total occupancy of the event
keeps the equalities of Eq.~(\ref{eq:fcum}), since the kept hadrons are a multinomial split of their
total, but violates the Gamma relation of Eq.~(\ref{eq:gamma3}). Let the efficiency fall linearly with
the number of hadrons produced in the three unit windows from $y_0=4$. The information then exceeds
the Gaussian value formed from the observed partial correlation once the efficiency at the mean of
that number lies more than $\NsOccLo$ percent below its value in an empty event at $k=3$, or $\NsOccHi$
percent at $k=1$. A displacement common to all hadrons of an event and independent of
$W$ makes the window means vary from event to event, and the counts are then no longer Poisson counts of one Gamma source.

The factorial cumulants of Sec.~\ref{sec:law} test necessary conditions for the window law, and they separate clusters of
hadrons, and pairs of hadrons straddling a window edge, from a source that is not Gamma. Clusters
raise the mean of the three within-window values of $F_{jl}$ above the mean of the three between-window
values, and pairs straddling a window edge raise $F_{12}$ and $F_{23}$ above $F_{13}$. A source that is not Gamma leaves Eq.~(\ref{eq:fcum})
intact and violates Eq.~(\ref{eq:gamma3}) unless its third cumulant equals $2\kappa_2^2$. The same
holds for a background of Poisson counts that are independent of $W$ and make up the same fraction of
the mean in every window. The equalities of Eq.~(\ref{eq:fcum}), and the corresponding equalities
at every order, are implied by one Poisson-counted source and do not establish it, since a
multinomial split of any total satisfies all of them. The between-window entries are not affected by
clusters that stay inside the window of their dipole. If every dipole gives $X\ge1$ hadrons,
independently of the other dipoles and with a law of $X$ that does not depend on $W$, and all of them land in the window where the dipole was produced, then
$F_{jl}=\mathrm{Var}\,W$ for $j\neq l$, while $F_{jj}$ rises by $\Ex[X(X-1)]/(\Ex[X]\,\lambda_j)$, and
the difference of the two sides of Eq.~(\ref{eq:gamma3}) becomes
$\kappa_{123}-2C_{12}C_{23}/\lambda_2=(\kappa_3-2(\mathrm{Var}\,W)^2)\,\lambda_1\lambda_2\lambda_3$
(Appendix~\ref{app:proof}). Giovannini and Ugoccioni give the two-hemisphere form of the statement on
the between-window entries, that clans, independently emitted groups of particles of common ancestry,
add no forward-backward correlation when each is confined to one hemisphere~\cite{GiovanniniUgoccioni}.

Wald tests of the second-order and the third-order equalities use the covariance of one
Poisson-counted source, with its moments estimated from the total count. Each Wald test forms the
differences that the equalities set to zero, called the contrasts, and refers their quadratic form,
weighted by the inverse of their estimated covariance, to a chi-square distribution. A third Wald test
compares the pooled cumulants with the Gamma relation, with the covariance of Eq.~(\ref{eq:nm}) at the fitted
means and shape, and a Bonferroni correction combines the three, rejecting when the smallest of the three
p-values is below $0.05/3$. We apply them to the unit windows
from $y_0=4$ at $k=1$, $1.84$, $3$ and $40$ and to three windows of equal mean $2$ at $k=3$. The shape $1.84$ is the value
$2h$ that Kutak and L\"ok\"os obtain from the entropy of proton-proton multiplicity
distributions~\cite{KutakLokos}. At the
nominal size of five percent, the chosen probability of rejecting a law that holds, the three tests and
their combination reject samples drawn from
Eq.~(\ref{eq:nm}) in $\NsSizeNullLo$ to $\NsSizeNullHi$ percent of cases. Each of these rates is pooled
over the configurations above and over samples of $10^4$, $10^5$ and $10^6$ events, with samples of
$10^7$ events added at $k=1.84$. The standard error of each rate is at most $\NsSizeSe$ percent. For the unit windows from $y_0=4$ the combination
rejects every sample of $10^4$ events drawn from clusters of two hadrons, each hadron displaced
independently in rapidity by a Gaussian of width $0.5$ about its dipole. The combination also rejects
every sample of $10^5$ events drawn from lognormal sources with the mean and variance of the Gamma.
The three tests check implications of the window law at second and third order, and passing them
does not establish the first inequality of Eq.~(\ref{eq:main}). A source that takes the values $2/3$
and $2$ with probabilities $3/4$ and $1/4$ has the first three cumulants of the Gamma of unit mean and shape $3$.
Since every normalized factorial cumulant of order $m$ is $\kappa_m(W)$, the contrasts of all three
tests vanish for this source, and their rejection rates do not grow with the sample size. For three
windows of equal mean $2$ its information left nevertheless exceeds $\Ig$ by $\NsTwoPointExcess$. The
information can also be estimated from the frequencies of the three counts and compared directly
with the Gaussian value. For the unit windows from $y_0=4$ this comparison holds its nominal size only
from $10^6$ events at $k=1$ and from $10^7$ at $k=3$.

\section{Beyond the $1{+}0$ model}\label{sec:beyond}

The derivation uses Poisson counting of one Gamma source, the form of Eq.~(\ref{eq:mixture}). With
the transverse dimensions restored, the splitting rate of a dipole depends on its size, and the sizes
present at a given rapidity change from event to event. We follow Mueller's cascade at large $N_c$,
fixed coupling and leading logarithmic accuracy~\cite{Mueller1994} in Monte Carlo simulations, with an
ultraviolet cutoff at $0.01\,r_0$ as in the program of Salam~\cite{OEDIPUS}, from one dipole of size
$r_0$ at $\alpha_s=0.15$ and $0.20$, with $10^7$ cascades for each setting. We count the splittings,
all of them and those whose smaller new dipole is at least $0.1\,r_0$ or $0.3\,r_0$, in three
consecutive unit windows of rapidity, with and without a cap on dipole sizes at $2r_0$. The lower edge
$y_0$ of the first window takes every integer value from $0$ to $7$ at $\alpha_s=0.15$ and from $0$ to
$5$ at $\alpha_s=0.20$, $\NtTriples$ window triples in all. In every triple the mean of the three
within-window values of $F_{jl}$ exceeds the mean of the three between-window values by at least
\NtExcessSigma{} standard deviations, so the counts are not Poisson counts of one source and the first
inequality of Eq.~(\ref{eq:main}) is not guaranteed. We estimate the information left from the
frequencies of the three counts with the entropy estimator of Grassberger~\cite{Grassberger2003}. An
estimate from finite frequencies is biased, and we measure the bias, for each window triple, on one sample of
the law of Eq.~(\ref{eq:nm}), whose information is known by direct summation, and subtract it. We call this
sample the bias control. It has the number of events of the cascade sample, its three window means, and $k$ equal to the inverse
of the normalized second factorial cumulant of the total count of the three windows. Where more distinct values of
$(n_1,n_2,n_3)$ occur in the cascade sample than in the corresponding sample of Eq.~(\ref{eq:nm}), the
size of the bias is multiplied by the ratio of the two numbers. The corrected estimate is taken to lie
above or below the Gaussian value only where it differs from that value by more than twice this size
plus four standard errors of the difference, which include the statistical error of the bias, and is
otherwise called undecided. This transfer of the bias to the cascade is
an empirical rule, with or without the rescaling, and the rescaled size of the bias is not proved to
bound the bias of the estimate for the cascade. The Gaussian value is formed in two ways that
coincide in the $1{+}0$ model, from the covariance matrix of the three counts and from the two
regression slopes of Eq.~(\ref{eq:reg}). With the cap no triple is found above either form. Without
the cap, $\NtAbove$ of the $\NtNoCapTriples$ triples lie above the covariance form, one of them with a
central value $\NtAboveRatio$ times that form, and none lies above the slope form. When the cutoff is
lowered to $0.001\,r_0$, with the count of all splittings replaced by the count of splittings whose
smaller new dipole is at least $0.01\,r_0$, the same $\NtAbove$ triples lie above the covariance form
and none lies above the slope form. At this cutoff two triples lie near the decision threshold. In
$\NtDrawsFurther$ further draws of the bias control, one of the $\NtAbove$ triples above is undecided in
$\NtDrawsUndec$, and a triple undecided in the draw used here lies above the covariance form in
$\NtDrawsAbove$. Hentschinski, Kutak, P{\l}aczek and Rohrmoser solve the complete
cascade with transverse dimensions at fixed coupling, with an infrared cutoff at the size of the
initial dipole, for the distribution of the total dipole number~\cite{HKPR}.

\section{Published data}\label{sec:data}

LHCb publishes the multiplicity distributions in five consecutive pseudorapidity intervals of width
$0.5$ with the means of the distributions truncated at twenty particles~\cite{LHCbmult}. The counted particles are prompt
charged particles with transverse momentum above $0.2$~GeV/$c$ and momentum above $2$~GeV/$c$, in
events with at least one such particle in $2.0<\eta<4.8$. For the three consecutive triples
the ratio $\lambda_1\lambda_3/\lambda_2^2$ is $\NumLHCbRatioOne$, $\NumLHCbRatioTwo$ and
$\NumLHCbRatioThree$, below unity, so for the window law at these means Eq.~(\ref{eq:sharp}) holds and
the ceiling applies. The joint distribution of the counts on one set of events is not published.

In the most central gold-gold collisions, STAR reports a correlation of $0.6$ between two
pseudorapidity intervals at fixed multiplicity in a reference interval five times as
wide~\cite{STARfb}. For two intervals placed symmetrically about the reference interval, as in this
measurement, Bzdak shows that the correlation equals the partial correlation of the two intervals
whenever the mean count of each is linear in the reference count, and quotes $0.58\pm0.06$~\cite{Bzdak}. The window
law has this linear regression, Eq.~(\ref{eq:reg}), and at that geometry, with the flat density STAR
reports, Eq.~(\ref{eq:rho}) gives $\rho=\lambda_1/(k+\lambda_1+\lambda_2)$, which stays below $1/6$ at
every $k$. The window law is therefore excluded in central gold-gold collisions at $\NsStarSigma$
standard deviations, and at about $\NsStarSigmaZdc$ if the same uncertainty is assigned to the value
close to one half that STAR finds with the centrality from the zero-degree
calorimeters~\cite{STARfb,Bzdak}.

The window law also fixes the normalized second factorial cumulant of one window. With $n$ the count in
that window it is $F=(\mathrm{Var}\,n-\langle n\rangle)/\langle n\rangle^2=\mathrm{Var}\,W=1/k$ in every
window fixed relative to the start of the cascade, and a cascade from one dipole gives $F=1$ when the
hadron of the initial dipole itself is not counted. Calucci and Treleani made the same point for a
negative binomial over the whole rapidity range from one source without correlation, whose shape is
then the same in every part of the range~\cite{CalucciTreleani}. H1 publishes the mean and the variance
of the charged multiplicity with transverse momentum above $150$~MeV and pseudorapidity
$|\eta_{\rm lab}|<1.6$ in the laboratory frame,
restricted to $0<\eta^*<4$ with $\eta^*$ the pseudorapidity in the hadronic center-of-mass frame, in
\NumHOneCells{} cells of photon virtuality and inelasticity~\cite{H1mult}. The mean and the variance give
$F$ between $\NtFMin$ and $\NtFMax$, so that the value one lies $\NtZLo$ to $\NtZHi$ standard deviations
above H1's value of $F$ in each cell, with the statistical, systematic and rounding uncertainties taken as independent
and the rounding uncertainty of each printed value taken as half a unit of its last digit. Clusters formed independently for
each dipole, with sizes that do not depend on the source $W$, add a positive term to $F$. A
displacement common to all hadrons of an event makes the window mean vary from event to event, and
this variation also raises $F$ when the displacement is independent of $W$ and of the positions of
the hadrons given $W$.
Within one kinematic cell the start of the cascade in
$\eta^*$ changes from event to event with the event kinematics, so the additive constant of
Sec.~\ref{sec:robust} is not common to the events of the cell, and the position in $\eta^*$ of the
edges of $|\eta_{\rm lab}|<1.6$ also changes from event to event. Each of the two acts as a
displacement common to all hadrons of an event, and both are set by the kinematics of the event. If
the kinematics of an event are fixed before its cascade develops, both displacements are independent
of $W$ and of the positions of the hadrons given $W$, and none of these effects then brings $F$ below
one. To lower $F$ to H1's values from one dipole, a Poisson background
independent of $W$ would have to
make up $\NtBgLo$ to $\NtBgHi$ percent of the window mean. Let an efficiency fall exponentially with the
number of hadrons in $|\eta_{\rm lab}|<1.6$, with H1's published mean there held as the mean after the
loss. In the cells where this efficiency can lower $F$ to H1's central value, the loss at the mean
occupancy would have to be $\NtEffLo$ to $\NtEffHi$ percent or more, and in the other \NtOccUnreach{} cells it
cannot reach that value at all. A requirement of at least one charged hadron in $|\eta_{\rm lab}|<1.6$
would also lower $F$, but H1's data carry no such requirement, since it would make the probability of no
charged hadron there zero, and H1 gives $\NtPzeroLo$ to $\NtPzeroHi$ percent~\cite{H1mult}. The window law contains no energy-momentum
conservation. In an earlier measurement H1 finds that the distribution of $n/\langle n\rangle$ widens
as the window narrows and relates this widening in part to the diminishing influence of global
conservation constraints~\cite{H1mult1996}. The H1 multiplicities are not described by a cascade from
one dipole counted as one hadron per dipole produced and none for the initial dipole, with independent
migration and efficiency, with the kinematics of each event fixed before its cascade develops, and
without such conservation.

\section{Relation to the entanglement entropy picture}\label{sec:interpret}

Kharzeev and Levin~\cite{KL} assume that the proton is in a pure state. For the partition into the
region probed and the rest, they identify the Schmidt basis with states of definite parton number.
Under both assumptions, the Shannon entropy of the parton multiplicity distribution equals the von
Neumann entropy of the reduced state of the region probed. Duan, Akkaya, Kovner and Skokov examine
that identification of the Schmidt basis and show
that for a complete set of number observables the von Neumann entropy of any state with the measured
distribution is at most the Shannon entropy of that distribution~\cite{DuanAKS}. The three counts
studied here are hadron multiplicities in disjoint rapidity windows, and the quantity bounded is a
classical conditional mutual information between them. It equals the quantum conditional mutual
information $S_{12}+S_{23}-S_{123}-S_{2}$, with $S_A$ the von Neumann entropy of the reduced state of
the windows in $A$, when the reduced state of the three windows is diagonal in the basis of definite counts
in each window. The assumptions of Kharzeev and Levin concern one partition and do not imply this
condition. For example, we take the pure state
$\sum_{n_1,n_2,n_3}\sqrt{P(n_1,n_2,n_3)}\,|n_1,n_2,n_3\rangle\,|e_N\rangle$, with $P$ the window law of
Eq.~(\ref{eq:nm}), $|n_1,n_2,n_3\rangle$ the state of the three windows with these counts and
$|e_N\rangle$ orthonormal states of the rest that depend only on the total count $N$. For the partition
into the three windows and the rest, its Schmidt vectors have a definite total count, one for each
value of $N$, so the state satisfies the assumptions of Kharzeev and Levin for this partition. By the
multinomial split of Sec.~\ref{sec:law}, the division of $n_1+n_2$ between the first two windows does
not depend on $n_3$, so each block of the reduced state of these two windows at fixed $n_1+n_2$ has
rank one, and likewise for the last two windows at fixed $n_2+n_3$. The reduced state of the three
windows has one rank-one block for each value of $N$, and that of the middle window is diagonal in its
count, since different values of $N$ go with orthogonal states of the rest. The quantum conditional
mutual information of this state is therefore
$H(n_1+n_2)+H(n_2+n_3)-H(n_1+n_2+n_3)-H(n_2)$, with $H$ the Shannon entropy of the distribution of the
count named, as in Eq.~(\ref{eq:cmidef}). For the unit windows of Sec.~\ref{sec:bound} from one dipole at $y_0=0$ it is
$\NtQcmiZero$, above $\tfrac12\ln(4/3)$, while the classical value is $\NumCmiShortZero$.

\section{Conclusions}

In the $1{+}0$ model of Mueller's dipole cascade~\cite{Mueller1995,KL}, with one counted particle for
each dipole produced in a window, the three counts are Poisson counts of one Gamma source. For these counts
the information left is below the Gaussian value of the partial correlation of the outer two at every
depth, width,
splitting rate and number of initial dipoles, Eq.~(\ref{eq:main}). The Gaussian value stays below the
ceiling $\tfrac12\ln(4/3)$ wherever the window means satisfy $\lambda_1\lambda_3\le\lambda_2^2$, as for
windows of one common width at a constant splitting rate, Eq.~(\ref{eq:bound}). The window law
keeps its form under the independent migration and loss of the counted particles, with new window
means, and its
factorial cumulants of second and third order test necessary conditions for it. For another source law of the same mean and variance the information can
exceed the Gaussian value, and in Monte Carlo simulations of the cascade with transverse dimensions at leading
logarithmic accuracy and fixed coupling, the counts are not Poisson counts of one source, so the bound is
a result of the $1{+}0$ model. In these simulations, with the empirical bias correction of Sec.~\ref{sec:beyond}, the information
left is estimated to exceed the Gaussian value formed
from the covariance matrix of the three counts in $\NtAbove$ of the $\NtTriples$ window triples, and
the Gaussian value formed from the two regression slopes in none. The simulations do not settle whether the
Gaussian value formed from the regression slopes bounds the information left in the cascade with
transverse dimensions.

\section*{Data and program availability}

Every computed value in the text is taken from the saved output of the programs that accompany this
paper as ancillary files. The program \texttt{make\_numbers.py} writes these values into the
manuscript and also checks the counts and bounds given in words. Configuration settings and values
quoted from the cited papers are stated as chosen or as published. The programs, their saved output
and the command that reproduces each are listed in the file \texttt{README.txt} of the ancillary
directory. The event samples of the Monte Carlo simulations are not included, and neither are several
files of per-sample results, five of them above 1~MB. The programs regenerate them from fixed seeds,
and \texttt{README.txt} lists them. Nor are the texts of the cited papers that some of the programs compare with included, and
\texttt{README.txt} names the source of each. Figure~\ref{fig:history} is one
realization drawn by a self-contained ROOT macro from a fixed random seed, and Fig.~\ref{fig:bound} is
drawn by ROOT from data files written by those programs.

\section*{Acknowledgments}

We thank Sergey Bondarenko for useful discussions. Appendix~\ref{app:ai} describes how large language
models were used in preparing this paper.

\appendix

\section{Derivations}\label{app:proof}

The Poisson representation of the cascade follows from the path likelihood. With $k$ dipoles at $y=0$
and a rate $\Delta(y)$ with integral $\Omega(y)$, the $i$-th production occurs at rate
$\Delta(t_i)(k+i-1)$, and the probability density of productions at $t_1<\dots<t_n$ in $[0,T]$ and
none else is
\begin{equation}
  \frac{\Gamma(k+n)}{\Gamma(k)}\prod_{i=1}^n\Delta(t_i)\,e^{\Omega(t_i)}\;e^{-(k+n)\Omega(T)} .
  \label{eq:path}
\end{equation}
A Poisson process of intensity $Wk\,\Delta(y)e^{\Omega(y)}$, mixed over $W$ Gamma of shape $k$ and unit
mean, gives the density of Eq.~(\ref{eq:path}) as well, since
$k^n\Ex[W^ne^{-kW(e^{\Omega(T)}-1)}]=\Gamma(k+n)\Gamma(k)^{-1}e^{-(k+n)\Omega(T)}$. The two laws
therefore coincide on every path. The window counts of this process are Eq.~(\ref{eq:mixture}) with
means $\lambda_j=k\,(e^{\Omega(y_j)}-e^{\Omega(y_{j-1})})$, which reduce to Eq.~(\ref{eq:means}) at
constant rate.

The function $\chi$ of Eq.~(\ref{eq:Jpp}) tends to zero as $z\to0$, rises, and stays below
$\tfrac12$, so $0<\chi<\tfrac12$. Its derivative $\chi'(z)=(2z+1)\ln(1+1/z)-2$ is positive precisely
when $\ln(1+u)>2u/(2+u)$ with $u=1/z$. The difference vanishes at $u=0$ and has derivative
$u^2/[(1+u)(2+u)^2]>0$. The expansion $\chi=\tfrac12-\tfrac1{6z}+\tfrac1{12z^2}-\dots$ gives the limit
$\tfrac12$. Equation~(\ref{eq:Jpp}) itself follows from two applications of
$\Ex[n\,f(n)]=\tfrac{c}{1+c}\,\Ex[(K+n)f(n+1)]$, the recursion $(n+1)P(n+1)=\tfrac{c}{1+c}(K+n)P(n)$
of the negative binomial of shape $K$ and mean $cK$. The same expansion of $\chi$ gives the approach to
$\Ig$. Since $\Ex[1/(K+n)]=1/[K(1+c)]+O(K^{-2})$, Eq.~(\ref{eq:Jpp}) gives
$\Theta''=1/[6K(1+c)^3]+O(K^{-2})$, so that $\Theta(c)=c^2/[12K(1+c)]$ up to a term linear in $c$ and
terms of order $K^{-2}$. The linear term cancels in Eq.~(\ref{eq:proof}), which gives
$\Ig-M(K)=\Xi/K+O(K^{-2})$.

Equation~(\ref{eq:sharp}) follows from Eq.~(\ref{eq:IG}), since
$4(1+c_1+c_3)-3(1+c_1)(1+c_3)=1+c_1+c_3-3c_1c_3$. It holds whenever $\lambda_1\lambda_3\le\lambda_2^2$.
Since $\lambda_1+\lambda_3\ge2\sqrt{\lambda_1\lambda_3}$, the condition written in means,
$3\lambda_1\lambda_3\le(k+\lambda_2)(k+\lambda_2+\lambda_1+\lambda_3)$, has a right-hand side at least
$(k+\lambda_2)(k+\lambda_2+2\sqrt{\lambda_1\lambda_3})$, which exceeds $3\lambda_1\lambda_3$ because
$k+\lambda_2>\lambda_2\ge\sqrt{\lambda_1\lambda_3}$.

The small-mean limit of Sec.~\ref{sec:bound} follows from an expansion of the information for Poisson
counts of one source $W$ of unit mean and variance $v>0$. Given $n_2+n_3$, the split between $n_2$ and $n_3$ is binomial and involves neither
$W$ nor $n_1$, so $\cmi=\mathcal{I}(\lambda_1,\lambda_2+\lambda_3)-\mathcal{I}(\lambda_1,\lambda_2)$,
with $\mathcal{I}(a,b)$ the mutual information of two counts that are Poisson of means $aW$ and $bW$
given $W$. This mutual information is a sum over the cells $(\nu_1,\nu_2)$ of the two counts of the
non-negative terms $q\,(t\ln t-t+1)$, with $p$ the joint probability of the cell, $q$ the product of
its marginal probabilities and $t=p/q$. As $a$ and $b$ tend to zero in fixed ratio, the cell $(1,1)$
has $p=ab\,\Ex[W^2e^{-(a+b)W}]$ and $q=ab\,\Ex[We^{-aW}]\,\Ex[We^{-bW}]$, equal to $ab(1+v)$ and $ab$
at leading order, so this cell contributes $ab\,\varphi(v)$. If $\Ex W^{2+d}$ is finite for some
$d>0$, which a finite third moment implies, all other cells together contribute at higher order in
$a$ and $b$. Hence $\cmi=\lambda_1\lambda_3\varphi(v)$ at leading order. At the same order
Eq.~(\ref{eq:IG}), with $k$ in the ratios $c_j$ replaced by $1/v$, gives $\Ig=\tfrac12
v^2\lambda_1\lambda_3$. Since $\varphi'(v)=\ln(1+v)<v$, the function $\varphi(v)$ lies below $v^2/2$ for
$v>0$, and the ratio tends to $2\varphi(v)/v^2<1$. The higher moment is needed only for the cells with
$\nu_1,\nu_2\ge1$ and $\nu_1+\nu_2\ge3$, and a finite variance alone does not suffice. The description
at the head of the ancillary program \texttt{small\_mean.py} derives the bound on each group of cells,
and the program evaluates the ratio at window means down to $10^{-6}$ for four source laws. The
ancillary program \texttt{counterexample.py} gives a source of finite variance with $\Ex W^{2+d}$
infinite for every $d>0$, for which the ratio grows without bound along a sequence of window means
tending to zero.

For the between-window identities of Sec.~\ref{sec:robust}, let each dipole produced in window $j$
give $X$ hadrons in the same window, with $\Ex[X]=m_1$ and $\Ex[X(X-1)]=m_2$, and let the dipole counts
be Poisson of means $\mu_jW$ given $W$. The hadron count $n_j$ is then compound Poisson given $W$, with
mean $m_1\mu_jW$, so the hadron means are $\lambda_j=m_1\mu_j$. Distinct windows receive the hadrons
of distinct dipoles, so given $W$ their counts are independent and every joint cumulant of distinct
windows comes from $W$ alone, $C_{jl}=\lambda_j\lambda_l\mathrm{Var}\,W$ for $j\neq l$ and
$\kappa_{123}=\lambda_1\lambda_2\lambda_3\,\kappa_3$. Hence $F_{jl}=\mathrm{Var}\,W$ for $j\neq l$ and
$\kappa_{123}-2C_{12}C_{23}/\lambda_2=(\kappa_3-2(\mathrm{Var}\,W)^2)\lambda_1\lambda_2\lambda_3$. Within
one window, $\Ex[n_j(n_j-1)\,|\,W]=m_2\mu_jW+(m_1\mu_jW)^2$, which gives $F_{jj}=\mathrm{Var}\,W+m_2/(m_1\lambda_j)$.

\section{Use of large language models}\label{app:ai}

The authors set every calculation and its configuration, fixed the working precision and the
stopping criterion of each one, and verified every derivation and every computed value reported here
by independent recomputation, with the values the accompanying programs compute generated from them
and not transcribed. Large language models were used in preparing this manuscript, specifically
Claude Opus 5, Claude Opus 5.5, Claude Sonnet 5 and Claude Fable 5.1 of Anthropic, GPT-5.6 Sol and
GPT-6 Astra of OpenAI, Gemini 3.1 Pro of Google, and Kimi K3 of Moonshot AI. On the authors' instructions they wrote the
programs that accompany this paper, which produce every computed value reported here and draw the
figures, they retrieved the cited papers, and they drafted and edited parts of the text. The figures
are drawn by ROOT from those programs, and no image-generating tool was used. Some of the
derivations and some of the numbers were additionally re-derived by a second model as a
cross-check. The authors reviewed and refined all model-generated text and are fully responsible for
the accuracy and the content of the paper. The models are not authors and made no contribution that
meets the criteria for authorship.

\end{document}